\documentclass[aps,prl,twocolumn,showpacs,preprintnumbers,amsmath,amssymb]{revtex4}
\usepackage{graphicx}% Include figure files \usepackage{bm}% bold
\usepackage{txfonts}

\begin{document}

%\preprint{APS/123-QED\\}

\title{Apparent violation of the Wiedemann-Franz law near a magnetic field tuned metal-antiferromagnetic quantum critical point}
\author{M. F. Smith}
 \email{mfsmith@physics.uq.edu.au}
\affiliation{%
University of Queensland, Department of Physics, 4072 Brisbane, Queensland, Australia}
\author{Ross H. McKenzie}
\affiliation{%
University of Queensland, Department of Physics, 4072 Brisbane, Queensland, Australia}
\date{\today}% It is always \today, today,

\begin{abstract}
The temperature dependence of the interlayer electrical and thermal resistivity in a layered metal are calculated for Fermi liquid quasiparticles which are scattered inelastically
by two-dimensional antiferromagnetic spin fluctuations.  Both resistivities have a linear temperature dependence over a broad temperature range. Extrapolations to zero temperature made from this linear-$T$ range give values that appear to violate the Wiedemann-Franz law.  However, below a low-temperature scale, which becomes small close to the critical point, a recovery of this law occurs. Our results describe recent measurements on CeCoIn$_5$ near a magnetic field-induced quantum phase transition.  Hence, the experiments do not necessarily imply a non-Fermi liquid
ground state.

\end{abstract}


\maketitle

Strongly correlated electron materials exhibit a subtle
competition between a range
of ground states including metallic, insulating,
 superconducting,
antiferromagnetic, and paramagnetic.\cite{dago05}
Often the metallic states are
distinctly different from the Fermi liquid
state characteristic of elemental metals.
Heavy fermion metals are particularly interesting
because they undergo quantum phase transitions, and
have non-Fermi liquid properties near
the quantum critical point.\cite{stew96,gegg08}
For example, the material family CeMIn$_5$ (where M=Co,Rh,Ir) can be tuned
through transitions associated
with antiferromagnetic or superconducting order
by varying magnetic field, pressure, or
chemical composition.
Understanding these systems has motivated significant
theoretical effort.\cite{mori95,si01,yang08}
In CeCoIn$_5$ at ambient pressure a quantum phase transition
between superconducting and metallic states
occurs as the magnetic field $H$
is tuned through a critical value $H_c$.\cite{petr01,pagl03,banc03}  Recent transport data\cite{tana07}, from the vicinity of this critical point, display an extraordinary violation of the fundamental Wiedemann-Franz (WF) law of metals, and have been interpreted\cite{cole07} as a possible signature of a non-Fermi liquid ground state of CeCoIn$_5$.

In this Letter we consider the WF law near a quantum-critical point with a goal to understand what  observable WF-violation (WFv) reveals about the electronic ground state.  The law states\cite{ash} that the electrical resistivity $\rho(T)$ is equal to the electronic thermal resistivity\cite{wnote} $w(T)$.  While it should not hold at finite $T$, since inelastic scattering may be important, the WF law must be obeyed by a Fermi liquid at $T=0$ where scattering is due to static defects.\cite{abrik}  An intriguing aspect of the CeCoIn$_5$ data is that $T\to 0$ intercepts of $\rho(T)$ and $w(T)$, extrapolated from the lowest observed $T$ of roughly 50 mK,  have a finite difference $w(T\to 0)-\rho(T\to 0)$ that increases as $H$ is tuned towards $H_c$.  This makes it appear that WFv might persist down to $T=0$, revealing a breakdown of the Fermi liquid ground state.  However, we show that a Fermi liquid subject to scattering by 2D critical spin fluctuations exhibits WFv at finite $T$, and in $T\to 0$ extrapolations made from above a low-temperature scale, while still obeying the WF law at $T=0$.  Since a Fermi liquid model captures the distinctive $H$ and $T$ dependence of the CeCoIn$_5$ data, these data\cite{tana07} do not necessarily imply a failure of the Fermi liquid picture.

We first place WFv within the context of transport phenomenology in CeCoIn$_5$.  The resistivity $\rho(T)$ is linear in $T$ below 40 K over a wide range of field $H$ and chemical doping, which could be an indication of quantum critical behavior\cite{rosc00} although its precise origin is unclear.\cite{pagl06}  For current along the stacked CeIn$_3$ planes, i.e., intralayer current, linear-$T$ $\rho(T)$ extends down to 5 K, roughly the same $T$ below which antiferromagnetic correlations appear\cite{curr03,stoc07}), before a downturn with decreasing $T$.  Below 1 K, intralayer $\rho(T)$ and $w(T)$ converge, suggesting that the WF law is obeyed for intralayer currents at sufficiently low $T$ (the WF law is also approximately obeyed above 5 K in intralayer data\cite{pagl06}).  Interlayer $\rho(T)$ is $T$-linear down to the lowest measurable $T$ with no trace of a downturn.  For $H$ well above $H_c$, $\rho(T)$ and $w(T)$ extrapolate to similar values at $T=0$. But as $H$ is decreased towards $H_c$, the interlayer thermal resistivity $w(T)$ undergoes a rigid upward shift that results in extrapolated $T\to 0$ WFv.  Since $w(T\to 0)>\rho(T\to 0)$, the WFv cannot be due to heat transport by neutral carriers\cite{onos07} but suggests instead that inelastic scattering contributes to the $T\to 0$ resistivities.

Since the $T\to 0$ WFv is less robust than $T$-linear $\rho(T)$, it can be plausibly attributed to a different mechanism.  WFv might result from scattering by inelastic antiferromagnetic spin fluctuations active at low temperatures, with $T$-linear $\rho(T)$ determined by another scattering process important over a wider temperature range\cite{yang08}.  Taking into account its experimental Fermi surface\cite{sett01,hall01}, we model CeCoIn$_5$ as a quasi-2D metal subject to strong two-dimensional antiferromagnetic spin fluctuations and study interlayer transport using a Boltzmann-equation description.  The approach aims to understand the low-temperature WFv in interlayer transport without addressing other unusual properties associated with larger temperature scales.

The relaxation rate of interlayer currents $\tau^{-1}=\tau_0^{-1}+\tilde{\tau}^{-1}$ is written as a sum of an isotropic, elastic part $\tau_0^{-1}$ and an anisotropic, inelastic part $\tilde{\tau}^{-1}=\tilde{\tau}^{-1}(\phi,\epsilon,T)$, coming from critical spin fluctuations.  At low-$T$, in the critical region, $\tilde{\tau}^{-1}$ will be active only near hot spots, i.e. near a pair of points on the 2D Fermi surface (of a single layer) that are connected by a spin-ordering wavevector ${\bf Q}$.  (To have finite interlayer current, the Fermi surface must have some modulation along the interlayer momentum $k_z$.  If spin fluctuations are spatially confined to a layer, and thus capable of imparting an arbitrary momentum transfer along $k_z$, then hot `spots' will actually be lines nearly-parallel to the $k_z$-axis on the quasi-2D Fermi surface.)  The inelastic scattering rate depends on the direction $\phi$ of electron momentum within a layer, on energy $\epsilon$ and temperature $T$.  The electrical and thermal resistivity can both be written, using $\eta^{-1}_0(T)\equiv \rho^{-1}(T)$ and $\eta_2(T)\equiv w^{-1}(T)$, as
\begin{equation}
\label{sig}
\eta_n^{-1}(T)=\eta_n^{-1}\int_{-\infty}^{\infty} dx x^n \bigg{(}\frac{-df_0}{dx}\bigg{)} \int_{-\bar{\phi}/2}^{\bar{\phi}/2} \frac{d\phi}{\bar{\phi}}\Lambda(\phi,x,T)
\end{equation}
where $\Lambda^{-1}(\phi,x,T)=1+\tau_0\tilde{\tau}^{-1}(\phi,x,T)$,
$\eta_n^{-1}=\rho_0^{-1}(3/\pi^2)^{n/2}$, $\rho_0$ the zero-temperature interlayer resistivity and $f_0(x)=(1+\mathrm{e}^x)^{-1}$ is the Fermi function and vertex corrections have been omitted (see Ref. \onlinecite{vc}).  Here, we assume symmetry-equivalent hot spots spaced by angle $\bar{\phi}$ with one at $\phi=0$ (this assumption is not crucial, and any other distribution of distinct hot spots gives similar results).  At $T=0$ we have $\tau^{-1}=\tau_0^{-1}$ and $\rho(0)=w(0)=\rho_0$, so the WF law is obeyed at sufficiently low $T$ in this Fermi liquid model.  But since our understanding of $T=0$ properties is based on measurements made at finite temperature, the effect of $\tilde{\tau}^{-1}$ on $T\to 0$ extrapolations of the model should be considered.

The rate of scattering of electrons by spin fluctuations\cite{rosc00} (the lowest-order electron self-energy with the spin susceptibility as the boson propagator) is
\begin{equation}
\label{tau}
\tilde{\tau}^{-1}(\phi,x,T)\!=\!\!2g_s^2\sum_{\bf k^{\prime}}\frac{f_0(x^{\prime})n_0(x-x^{\prime})}{f_0(x)}\chi ''_{\tiny{\bf k}-{\bf k^{\prime}}}(k_BT[x-x^{\prime}])
\end{equation}
where the ${\bf k}^{\prime}$ sum is done in the usual way as an integral over linearized band energy $x^{\prime}=\epsilon_{{\bf k}^{\prime}}/k_BT$ and position on the Fermi surface $\phi^{\prime},k^{\prime}_z$, with ${\bf k}-{\bf k}^{\prime}$ dependent only on $\phi,\phi^{\prime}$. $n_0(x)=(\mathrm{e}^x-1)^{-1}$ is a Bose function and $\chi''_{\bf q}(\omega)$ is the imaginary part of the spin susceptibility,\cite{mori95,pine95,sach}
\begin{equation}
\label{chi}
(k_BT_2)^{-1}\chi^{-1}_{\bf q}(\omega)=-i\frac{\omega}{k_BT_2}+\omega_{\bf q}+(k_f\xi)^{-2},
\end{equation}
the energy scale $k_BT_2$ is the width of the spin-fluctuation spectral function at typical ${\bf q}$ (it is proportional to the parameter $T_0$ of Ref. \onlinecite{mori95}) and $g_s$ a coupling constant .  The factor $\omega_{\bf q}k_f^2=({\bf q}\pm{\bf Q})^2$ for $|{\bf q}\pm {\bf Q}|<<k_f$, while for $|{\bf q}\pm {\bf Q}|\approx k_f$, it is roughly ${\bf q}$-independent.  The spin fluctuations are assumed two-dimensional (the 3D case is discussed below) so $\chi_{\bf q}(\omega)$ is independent of ${\bf q}_z$.  Also, we take for the spin-correlation length\cite{sach} at $T<<T_2$
\begin{equation}
\label{xieq}
(\xi q_0)^{-2}=r+cT/T_2
\end{equation}
where $c$ is a constant of order unity and $r$, the tuning parameter, measures the proximity to the quantum critical point . For the magnetic field-tuned quantum critical point of interest $r$ depends on $H$ and vanishes at $H=H_C$.  In the expressions above we have ignored logarithmic corrections, associated with the system being in its upper critical dimension\cite{sach}.

\begin{figure}
\begin{center}
\includegraphics[width=2.4 in, height=2.0 in]{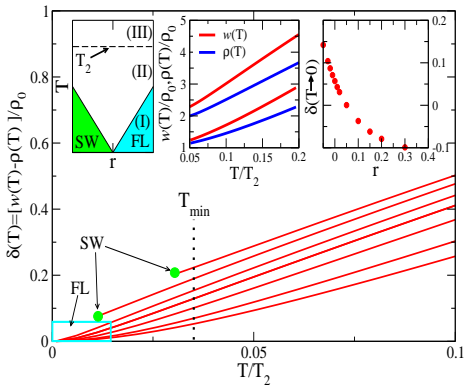}
\end{center}
\caption{\label{delplot} Wiedemann-Franz law violation near quantum critical point.  Left inset: Phase diagram showing critical point between a Fermi liquid (FL) and spin density wave (SW) with tuning parameter $r$.  (I)-(III) denote different temperature regimes discussed in text.  Center inset: The thermal $w(T)$ and electrical $\rho(T)$ resistivity close to the critical point ($r=-0.03$, upper curves) and further from it ($r=0.3$, lower curves).  Main panel:  The difference $\delta(T)\equiv [w(T)-\rho(T)]/\rho_0$ in resistivities for $r=0.3,0.1,0.05,0.03,0.01,-0.01,-0.03$ from bottom to top.  The supposed minimum measurable temperature is $T_{\mathrm{min}}$ (too large to see SW or FL states for $|r|<<1$).  The Wiedemann-Franz law $\delta(0)=0$ would appear to be violated based on extrapolations made from above $T_{\mathrm{min}}$.  Right inset: $T=0$ intercepts of $\delta(T)$ obtained from extrapolations from $T>T_{\mathrm{min}}$, which increase as $r$ is decreased.}
\end{figure}

We discuss three temperature regimes for the spin fluctuations (always remaining close to the critical point where $|r|<<1$), which are indicated respectively as I, II and III in the upper-left inset of Fig. \ref{delplot}.  At $T<<T_2$ only spin fluctuations close to an antiferromagnetic ${\bf Q}$ are thermally excited so only electrons near hot spots encounter inelastic scattering.  In the low-temperature region (I), defined by $T<<rT_2<<T_2$ (with $r\geq 0)$, the correlation length $\xi$ is determined by the tuning parameter, $r$ in Eq. \ref{xieq}, with temperature giving a weak correction.  This may be distinguished from an intermediate temperature regime (II), $rT_2<<T<<T_2$, in which this situation is reversed.  At high temperatures (III), $T>T_2$, all spin fluctuations are thermally excited so there are no hot spots and $\tilde{\tau}^{-1}$ is independent of $\phi$.

Another significant temperature scale is that at which the inelastic scattering rate becomes comparable to $\tau_0^{-1}$ (this is clarified below in the discussion of orbital effects of the magnetic field).  According to Eq. \ref{tau}, the strength of inelastic scattering is characterized by a temperature scale $T_1=(\Gamma\pi v_f c_L k_f |\sin\psi|\bar{\phi}\tau_0^{-1})(\Omega k_Bg_s^2 q_0^2)^{-1}$, where $\Omega$ is the sample volume, $c_L$ the $c$-axis lattice spacing and both the Fermi velocity $v_f$ and $\psi$, the angle between velocities ${\bf v_{\bf k}}$ and ${\bf v_{\bf k+Q}}$, are evaluated at a hot spot.

{\it Low-$T$ regime (I)} When $T<<rT_2$ we have
\begin{equation}
\label{fl}
\frac{\rho(T)}{\rho_0}=1+\bigg{(}\frac{\pi T^2}{3rT_1T_2}\bigg{)}\;,\;\frac{w(T)}{\rho_0}=1+\frac{9}{5}\bigg{(}\frac{\pi T^2}{3rT_1T_2}\bigg{)}.
\end{equation}
In regime (I) inelastic scattering becomes comparable to impurity scattering when $T\gtrsim \sqrt{rT_2T_1}$.  The WF law would be obeyed by $T=0$ extrapolations made from regime (I) but, close to the quantum critical point, this regime will be limited to inaccessibly low temperatures.

{\it Intermediate-$T$ regime (II)}  For $rT_2<<T<<T_2$,
\begin{equation}
\label{linT}
\frac{\rho(T)}{\rho_0}=1-a_0\frac{rT_2}{T_1}+b_0\frac{T}{T_1}\;,\;\frac{w(T)}{\rho_0}=1-a_2\frac{rT_2}{T_1}+b_2\frac{T}{T_1}
\end{equation}
where $a_n$,$b_n$ are positive numbers with the $n=2$ terms being larger.  $a_n$ are given by $a_n=(3/\pi^2)^{n/2}\int_{-\infty}^{\infty} dx (-df_0/dx) a(x) x^n$ and $b_n$ is the same with $a(x)$ replaced by $b(x)$ where $a(x)=\pi^{-1}\int_{-\infty}^{\infty} dx^{\prime} [f_0(x^{\prime}-x)+n_0(x^{\prime})]x^{\prime}(x^{\prime \;\;2}+c^2)^{-1}$, $b(x)=\pi^{-1}\int_{-\infty}^{\infty} dx^{\prime} [f_0(x^{\prime}-x)+n_0(x^{\prime})]\mathrm{sgn}x^{\prime}[\frac{\pi}{2}-\mathrm{atan}(\frac{c}{|x^{\prime}|})]$.
In regime (II) the inelastic scattering rate exceeds $\tau_0^{-1}$ once $T\gtrsim T_1$.  Extrapolations to $T=0$ from regime (II) violate the WF law.  This is because the $T=0$ intercepts of $\rho(T)$ and $w(T)$ are not due only to impurity scattering, they also include an inelastic contribution coming from the $r$-linear term in $\xi^{-2}$, Eq. \ref{xieq}.  (WFv results from inelastic electron-electron scattering, as studied rigorously in the context of disordered metals.\cite{cate04})

In Fig. \ref{delplot}, $\delta(T)\equiv [w(T)-\rho(T)]/\rho_0$ is plotted and extrapolated to $T=0$ from fits made above an arbitrary $T_\mathrm{min}$ (supposed to be the lowest measurable $T$) for several values of the tuning parameter $r$.  For small $|r|$, $T_{\mathrm{min}}$ is in regime (II), so $\delta(T)$ is linear in $T$ with an $r$-independent slope and a non-zero extrapolated $T=0$ intercept that increases as $r$ is decreased.  Extrapolations from above $T_{min}$ suggest $T\to 0$ violation of the WF law.  The low-$T$ recovery of the WF law, which occurs for $r\geq 0$, is unobservable.  For $r<0$, the model fails when $T$ is decreased to values comparable to $|r|T_2$ (i.e. it breaks down as spin-order is approached) and says nothing about $T\to 0$ WF behavior for $r<0$.  [Negative $r$ values are allowed in the model\cite{sach} within regime (II).]

The results in Fig. \ref{delplot} capture much of the low-$T$ behavior observed in the interlayer transport of CeCoIn$_5$.  Tanatar {\it et al}\cite{tana07} measured $\delta(T)\propto T$ with a non-zero intercept $\delta(T\to 0)$.   With decreasing field, $\delta(T)$ underwent a rigid upward shift and $\delta(T\to 0)$ increased from slightly negative values at high fields to positive values close to $H_c$.   We can make a semi-quantitative comparison with this data: the measured $\delta(T\to 0)$ decreases by 0.2 as the field goes from 5.3 to 6 T.  Using $r=H/H_c-1$, taking the constant $c=1$, and associating the measured field-dependence of $\delta(T\to 0)$ near $H_c$ with its predicted linear dependence on $r$, we obtain $T_2/T_1\approx 10$.  $T_1\approx 400$ mK is estimated independently from the slope of $\delta(T)$.  So $T_2\approx 4$ K, which is consistent with the value from neutron scattering\cite{stoc07} and with the temperature below which WFv begins and linear $T$-resistivity ends in intralayer transport\cite{pagl06}.  $T_2/T_1=10$ and $c=1$ are also used in Fig. \ref{delplot} and the plots may be compared to those\cite{plotnote} of data in Ref. \onlinecite{tana07}.  The $r$ values in Fig. \ref{delplot} correspond to $H$ ranging from 0.2 T below $H_c$ to 3.5 T above it, roughly consistent with a range used to fit the $T$-dependence of specific heat with the same model.\cite{banc03}  Negative $r$ values in CeCoIn$_5$ would imply that the $T=0$ critical field for the magnetic-paramagnetic transition $H_c$ is larger than the superconducting transition field at the lowest measured temperatures.  In the $r\gtrsim 0$ curves of Fig. \ref{delplot}, extrapolations made from $T$ as low as $0.01 T_2\approx 40$ mK would indicate WF violation, missing its low-$T$ recovery.

{\it High-$T$ regime (III)}  If we crudely extend the model to $T > T_2$, then it predicts isotropic, $T$-linear scattering.  (This assumes that $\chi(\omega=0)$ does not decrease as fast at $1/T$ so $\chi''(\omega)$, Eq. \ref{chi}, restricts $\omega$ integrals to $\omega<<k_BT$, which should be a reasonable approximation for $T\gtrsim T_2$.)  We then have
$\rho(T)=\rho_0(1+a_0^{\prime}T/T_1)$, and $\delta(T)\approx T_2^2/(\pi^2 T T_1)$
where $a_0^{\prime}\approx 1$ is a constant.

\begin{figure}
\begin{center}
\includegraphics[width=2.16 in, height=1.8 in]{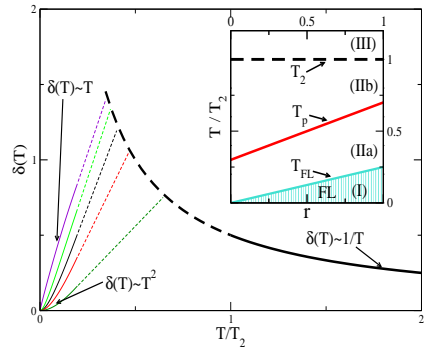}
\end{center}
\caption{\label{hiTplot} Qualitative behavior of $\delta(T)$ over a wide $T$-range.  Main Panel: Solid curves at low $T$ are plots of $\delta(T)$, as shown in Fig. \ref{delplot}, for decreasing $r$ between $r=1$ and $r=0$ from bottom to top.  The single solid curve at high $T$ is the ($r$-independent) approximate result, and dashed curves are extrapolations into the crossover regime. As $r$ is decreased, the peak in $\delta(T)$ narrows and shifts to lower $T$ but the peak-temperature $T_p$ remains finite at $r=0$.  Inset: The $r$-dependent temperature scales associated with $\delta(T)$.  The peak temperature $T_p$, lies between the Fermi liquid temperature $T_{\mathrm{FL}}$ and spin-fluctuation temperature $T_2$.}
\end{figure}

Fig. \ref{hiTplot} is an approximate plot of $\delta(T)$ over a wide temperature range: $\delta(T)$ increases with $T$ before peaking at a temperature $T_p$ and falling off like $1/T$.  This behavior is analogous to what is seen for electron scattering from phonons\cite{abrik} (with $T_2$ playing the role of the Debye temperature) where WFv is small both at low-$T$, where few phonons are excited, and high-$T$, where the phonon energy is small compared to thermal electron energy so scattering is elastic.  As $r$ is decreased, the peak in $\delta(T)$ narrows and its position $T_p$ shifts to lower $T$ but $T_p$ does not tend towards zero.  Rather, $T_p/T_2\approx (T_1/T_2)^{1/2}$ at $r=0$.  Similar behavior is seen in $\delta(T)$ data for intralayer transport\cite{pagl06,tana07} in CeCoIn$_5$, though in the latter data the resistivity is not $T$-linear over the temperature range in which the peak occurs.  

The above assumed 2D fluctuations as suggested by NMR data on CeCoIn$_5$\cite{curr03}.  Recent neutron data\cite{stoc07} see only weak spin anisotropy and a dimensional crossover below 1K (3D at low $T$) has been reported.\cite{dona07}  For 3D fluctuations (assuming hot-spots still exist), we write $\omega_{\bf q}q_0^2=({\bf q}_{\parallel}-{\bf Q}_{\parallel})^2+\alpha^2(q_z-Q_z)^2$ where ${\bf q}_{\parallel}$ is in the layer and $\alpha^2$ a measure of anisotropy, and take\cite{sach} $(\xi q_0)^{-2}=1+c(T/T_2)^{3/2}$.  We find $\rho(T)/\rho_0=a_{3D}[T^{5/4}/(T_1T_2^{1/4})-(r/2c)T^{-1/4}/(T_1T_2^{-1/4})]$ and $w(T)=(9/5) \rho(T)$ with $a_{3D}\approx 1.05/(\alpha c^{1/2})$.    So resistivities vary as $T^{5/4}$ and the field-dependent term is proportional to $T^{-1/4}$. A 3D-2D crossover would be difficult to observe in transport given the slight change in temperature power laws but careful analysis of the $T$-dependent WFv could reveal the dimensionality of the scatterer.

Quantum oscillations are seen\cite{goh08,sett01} in CeCoIn$_5$ for magnetic fields $H$ as low as $3 H_c$, so orbital effects of the field might be important near $H_c$. We have studied these using a Boltzmann-equation for interlayer transport with a field along $k_z$.  At $T<<T_2$, near $H_c$, hot-spot scattering is important since it reduces electron density in cold regions that follow hot spots, in the sense of cyclotron motion. One can solve the Boltzmann equation in cold regions, treating the density at the hot spot as a boundary condition obtained by solving the equation within the narrow range, say $-\delta\phi/2<\phi<\delta\phi/2$ where $\tilde{\tau}^{-1}(\phi)$ operates.  Eqs. \ref {fl} and \ref{linT} are valid if the cyclotron frequency $\omega_C\equiv eHv_f/k_f$ satisfies $\omega_C>\int _{-\delta\phi/2}^{\delta\phi/2} \tilde{\tau}^{-1}(\phi)$.  This is why $T_1$, the temperature at which the $\phi$-averaged $\tilde{\tau}^{-1}(\phi)$ equals $\tau_0^{-1}$, is a relevant scale.  The much lower $T$ at which $\tilde{\tau}^{-1}(0)=\tau_0^{-1}$ should not matter near $H_c$.  The field helps prevent the onset of an anisotropic electron distribution\cite{rosc00}) as $T$ is increased.

In summary, the field-tuned Wiedemann-Franz violation seen in extrapolated linear intercepts of interlayer resistivity in CeCoIn$_5$ is explained within a Fermi liquid picture in which two-dimensional antiferromagnetic spin fluctuations are the main source of inelastic scattering.  This picture naturally produces multiple temperature scales associated with transport and shows how these scales vary near the critical point.  

This work was supported by Australian Research Council Discovery Project DP0710617.  We thank B. J. Powell, S. Olsen, J. Merino, J. Paglione, N. Hussey and M. A. Tanatar for helpful feedback.


\begin{thebibliography}{99}
\bibitem{wnote} $w(T)$ is related to the thermal conductivity $\kappa(T)$ by $w^{-1}(T)=\kappa(T)/(L_0T)$ where $L_0=k_B^2\pi^2/(3e^2)$.
\bibitem{vc} For a quasi-2D system, vertex corrections to interlayer transport can be neglected to lowest order in the interlayer coupling\cite{mahan,smit08}.  Vertex corrections might be more important for intralayer transport, though they do not change the qualitative properties of $\rho(T)$ for scattering from impurities and antiferromagnetic spin fluctuations\cite{rosc00,pine95}.
\end{thebibliography}
\end{document}